\documentclass[
preprint,
amsmath,amssymb,
aps,
]{revtex4-2}

\usepackage{graphicx}
\usepackage{dcolumn}
\usepackage{bm}
\usepackage{hyperref}
\usepackage[dvipsnames]{xcolor}
\usepackage{comment}
\usepackage{multirow}
\usepackage{mathtools}
\usepackage{xcolor}
\usepackage{bigstrut,slashbox}
\usepackage{booktabs}
\usepackage{array}

\newcommand{\be}{\begin{equation}}
\newcommand{\ee}{\end{equation}}
\newcommand{\bea}{\begin{eqnarray}}
\newcommand{\eea}{\end{eqnarray}}
\newcommand{\beq}{\begin{equation}}
\newcommand{\eeq}{\end{equation}}
\def\bsp#1\esp{\begin{split}#1\end{split}}
\def\bal#1\eal{\begin{align}#1\end{align}}

\newcommand\rL   {\ensuremath{\mathrm{L}}}

\newcommand\MeV  {\ensuremath{\mathrm{MeV}}}
\newcommand\GeV  {\ensuremath{\mathrm{GeV}}}
\newcommand\TeV  {\ensuremath{\mathrm{TeV}}}
\newcommand\rc   {\ensuremath{\mathrm{c}}}

\newcommand\ri   {\ensuremath{\mathrm{i}}}

\newcommand\rs   {\ensuremath{\mathrm{s}}}

\newcommand\gL   {\ensuremath{g_\mathrm{L}}}
\newcommand\gy   {\ensuremath{g_{y}}}
\newcommand\gz   {g_{z}}
\newcommand\gZ   {\ensuremath{g_{Z^0}}}

\newcommand\cL   {\ensuremath{\mathcal{L}}}
\newcommand\GF {\ensuremath{ G_{\rm F}}}
\newcommand\cw {\ensuremath{ \rc_W}}
\newcommand\sw {\ensuremath{ \rs_W}}
\newcommand\cz {\ensuremath{ \rc_Z}}
\newcommand\sz {\ensuremath{ \rs_Z}}

\newcommand\sss {\ensuremath{ \rs_S}}
\newcommand\tanb {\ensuremath{ \tan\beta}}
\newcommand\msbar {\ensuremath{\overline{\text{MS}}}~}

\newcommand\Zb {\ensuremath{ Z^{(0)}}}
\newcommand\Zpb{\ensuremath{ Z^{\prime(0)}}}
\makeatletter
\newcommand{\undersim}[1]{\mathrel{\mathpalette\@undersim{#1}}}
\newcommand{\@undersim}[2]{%
  \vcenter{%
    \ialign{%
      ##\cr
      $\m@th#1#2$\cr
      \noalign{\nointerlineskip\kern.2ex}
      $\m@th#1\sim$\cr
      \noalign{\kern-.4ex}
    }%
  }%
}

\makeatother

\begin{document}

\title{Precise prediction for the mass of the $W$ boson in gauged U(1) extensions of the standard model}

\author{Zolt\'an P\'eli}
\email{zoltan.peli@ttk.elte.hu}
\affiliation{Institute for Theoretical Physics, ELTE E\"otv\"os Lor\'and University,
P\'azm\'any P\'eter s\'et\'any 1/A, 1117 Budapest, Hungary
}
\author{Zolt\'an Tr\'ocs\'anyi}
\email{zoltan.trocsanyi@cern.ch}
\affiliation{Institute for Theoretical Physics, ELTE E\"otv\"os Lor\'and University,
P\'azm\'any P\'eter s\'et\'any 1/A, 1117 Budapest, Hungary
%, and\\
%Institute of Physics, Faculty of Science and Technology, University of Debrecen,
%Bem ter 18/A, 4026 Debrecen, Hungary
}
\date{\today}
\begin{abstract}
We present the one-loop radiative corrections to the muon decay in U(1)$_z$ 
extensions of the standard model. We compute the mass of the $W$ boson using those corrections and compare it to an approximation of the 
complete one-loop prediction implemented in automated computational tools. We point out that the 
truncation of the complete formulas become unreliable if the mass of the $Z'$ boson, 
corresponding to the new U(1)$_z$ gauge group is larger than about 1\,TeV.
\end{abstract}
\maketitle

The measurements for extracting the mass of the $W$ boson at hadron colliders 
\cite{CDF:2012gpf,D0:2012kms,ATLAS:2017rzl,LHCb:2021bjt} are steadily improving,
and its precision is approaching the per myriad level \cite{CDF:2022hxs,ATLAS:2023fsi}.
The caveat is that the last two results differ quite significantly, and there is 
a vigorous research activity to find the origin of this disagreement.
The current combined world average of the Particle Data Group \cite{Workman:2022ynf}
$M_W = (80377\pm 12)\,\MeV$ does not include the CDF 2022 result. 

On the theoretical side the SM prediction has also reached a similar precision for $M_W$ by 
including the one-, two-loop and leading three-loop quantum corrections in 
perturbation theory \cite{Awramik:2003rn,Degrassi:2014sxa}. These experimental
and theoretical advances has elevated $M_W$ to a prime precision parameter of 
the standard model (SM), which any extension of the SM must respect. Currently, 
there is a $2\sigma$ discrepancy between the theoretical prediction of the SM 
and the world average for $M_W$, which does not warrant any new physics effect,
but one at least expects that any physics beyond the standard model (BSM) should
not worsen the agreement between the measured value and the theoretical prediction.
Thus we assume that the potential new physics contributions to $M_W$ must lie within the 
difference of the experimental value and the SM prediction and the corrections 
stemming from new physics must be determined with similar precision as the SM value.

The mass of the $W$ boson can best inferred from the muon decay width. In this letter
we compute the complete one-loop radiative corrections to the muon decay process, 
hence to $M_W$, for the first time in a specific class of extensions of the SM. 
We consider models where the SM gauge group supplemented by an additional U(1)$_z$ 
gauge symmetry and the particle spectrum includes a complex scalar field $\chi$ 
that is neutral under the standard model gauge interactions, but contributes to 
the mass of the neutral gauge bosons through spontaneous symmetry breaking (SSB). 
U(1) extensions of the SM are popular for in spite of being relatively simple, 
they can explain a multiple of BSM phenomena \cite{
Ma:1997nq,Ma:2001md,Nelson:2007yq,Adhikari:2008uc,Heeck:2010pg,Bhatia:2017tgo,Asai:2019ciz,Borah:2020swo}.

The specific example we have in mind is the superweak extension of the standard model 
(SWSM) \cite{Trocsanyi:2018bkm}, although different charge assignments are also possible,
and our formulae do not depend on the choice explicitly.
The SWSM  contains also three generations of SM sterile right handed neutrinos that
are clearly necessary for the cancellation of gauge and gravity anomalies 
and to explain the origin of neutrino masses. We do not include 
their effect here to simplify the parameter dependence in the numerical
analysis, but it can be seamlessly integrated into our complete 
one-loop prediction.

The Lagrangian of the scalar fields contains a potential energy with quadratic 
and quartic terms such that non-vanishing vacuum expectation value (VEV) $v$ of the 
Brout-Englert-Higgs (BEH) field breaks the usual SU(2)$_\rL\otimes$U(1)$_Y$ symmetry, 
while the VEV $w$ of the $\chi$ breaks the U(1)$_z$ symmetry via spontaneous symmetry 
breaking (SSB).
 
In addition to the appearance of the massive scalar bosons, the SSB generates mass 
terms also for the gauge bosons 
\be 
\bsp
\cL_{\text{M}}^{\text{VB}} = 
\frac{v^2}{2}&\biggl[
 \frac{\gL^2}{2}W_\mu^{+}W^{-\mu} + \frac{g_z^2}{2} \tan^2\beta \:B'_\mu B'^\mu
\\&
+ \frac{1}{4}\Big(\gy B_\mu + \bigl( g_z-g_{yz}\bigr)B'_\mu - \gL W_\mu^{3}\Big)^2 \biggr]
\,,
\label{eq:lagrangian-gaugemass}
\esp
\ee
where $\tan\beta = w/v$, $\gL$,  $\gy$ and $g_z$ are the SU(2)$_\rL$, U(1)$_Y$ and 
U(1)$_z$ couplings, while the mixing coupling $g_{yz}$ parametrizes the kinetic mixing 
between the $B_\mu$ and $B'_\mu$ fields \cite{Iwamoto:2021fup}. The fields 
$W^{\pm}_\mu = \bigl(W_\mu^{1}\pm \ri W_\mu^{2}\bigr)/\sqrt{2}$ are the charged, while 
the neutral gauge eigenstates are $B_\mu$, $B'_\mu$ (belonging to the U(1)$_Y$
and U(1)$_z$ symmetries) and $W^{3}_\mu$. The latter fields are related to 
the neutral mass eigenstates $A_\mu$, $Z_\mu$ and $Z^{\prime}_\mu$ via two rotations 
\beq
\left(\begin{array}{c}
    B_\mu \\
    W^{3}_\mu \\
    B'_\mu
\end{array}\right) =
\left( \begin{array}{ccc}
\cw & -\sw & 0 \\
\sw &  \cw & 0 \\
  0 &  0   &  1 \end{array}
\right)
\left( \begin{array}{ccc}
  1 &   0 &  0   \\
  0 & \cz & -\sz \\
  0 & \sz &  \cz \end{array}
\right)
\left(\begin{array}{c}
    A_\mu \\
    Z_\mu \\
    Z^{\prime}_\mu
\end{array}\right)
\label{eq:gauge-rotation}
\eeq
where we introduced the abbreviations ${\rm c}_X = \cos\theta_X$ and ${\rm s}_X = \sin\theta_X$ for mixing angles. The Weinberg angle $\theta_{\rm W}$ is defined as 
\be
\sw = \frac{g_y}{\gZ}\,,\text{~~with the abbreviation~} \gZ^2 = g_y^2+\gL^2\,,
\ee
so $e = \gL\sw$ where $\gL$ is the SU(2) gauge coupling and $e$ is 
the elementary charge. The $Z-Z'$ mixing angle $\theta_Z \in(-\pi/4,\pi/4)$ is defined 
implicitly in terms of effective couplings
\be
 \kappa = \frac{2 z_\phi \gz - g_{yz}}{\gZ }
\text{~~and~~}
\tau = \frac{2\gz}{\gZ}\tanb 
%\tau = \frac{2 z_\phi \gz}{\gZ}\tanb 
 \label{eq:kappa-tau}
\ee
as
\be
\tan(2 \theta_Z) = -\frac{2 \kappa}{1 - \kappa^2 - \tau^2}\,,
\label{eq:tZ}
\ee
with $z_\phi$ being the $z$-charge of the BEH scalar.
Then the masses of the gauge bosons are 
\be
\bsp
M_W  = \frac{1}{2}g_{\rm L} v 
\,,\quad
M_Z = \frac{M_W}{\cw} \sqrt{R(\cz,\sz)}
\,,\\
M_{Z'} = \frac{M_W}{\cw} \sqrt{R(\sz,-\cz)}
\,,\qquad~
\label{eq:vectormasses}
\esp
\ee
with $R(x,y) = \bigl(x - \kappa y\bigr)^2 + \bigl(\tau y\bigr)^2$. The free parameters 
from the extended gauge sector are either the Lagrangian couplings  $(g_z,\:g_{yz})$ 
or the effective couplings $(\kappa,\:\tau)$. The latter can be expressed as 
functions of experimentally more accessible parameters $M_{Z'}$ and $\theta_Z$ as
\begin{equation}\label{eq:mzpthetaz}
    \kappa = -\cz \sz ~\frac{M_Z^2 - M_{Z'}^2}{ \cz^2 M_Z^2 + \sz^2 M_{Z'}^2}
\,,\;
    \tau = \frac{M_Z M_{Z'}}{ \cz^2 M_Z^2 + \sz^2 M_{Z'}^2}
    \,.
\end{equation}
The well known SM relationship between the $W$ and $Z$ boson masses is modified at tree level to
\be\label{eq:mwmzmzp}
\frac{M_W^2}{\cw^2} = \cz^2 M_Z^2 + \sz^2 M_{Z'}^2.
\ee
This formula coincides with the one obtained later in Ref.~\cite{bento2023treelevel} using sum rules for tree level unitarity in U(1) extensions.

The mass of the $W$ boson with small theoretical uncertainty can be extracted from the muon 
decay width. The order $O(\alpha)$ corrections to the muon decay process in the SM can be
summarized by properly modifying the tree-level relation among the Fermi coupling $G_F$, 
the fine structure constant $\alpha$ and the mass of the $W$ boson, which was first derived 
in Ref.~\cite{Sirlin:1980nh}, 
\begin{equation}
    \frac{G_F}{\sqrt{2}}= \frac{\pi \alpha}{2 M_W^2 \Big(1-\cw^2\Big)}\bigl[ 1- \Delta r\bigr]^{-1}
    \label{eq:deltar-def}
\end{equation}
where the parameter $\Delta r$ collects the radiative corrections that enter electroweak 
precision observables, as well as it is used to express $M_W-M_Z$ interdependence. 
In a U(1)$_z$ extension, the relation among the masses of the gauge bosons including the radiative corrections follows from Eqs.~\eqref{eq:mwmzmzp} and \eqref{eq:deltar-def} as
\be 
\bsp
M_W^2 = &\frac{\cz^2 M_Z^2 + \sz^2 M_{Z'}^2}{2}
\\&\times
\left(1+ \sqrt{1- \frac{4 \pi\alpha/\bigl(\sqrt{2}\GF\bigr)}{\cz^2 M_Z^2 + \sz^2 M_{Z'}^2}\frac{1}{1-\Delta r}} \right).
\esp
\ee

We can classify the quantum corrections to the tree level amplitude for the muon decay 
process into three categories: 
(i) the renormalization of the SU(2) gauge coupling $\gL$ collected into the counterterm 
$\delta \gL$,
(ii) loop corrections to the $W$ boson propagator and 
(iii) contribution of the vertex and box loop diagrams $\delta_{\text{BV}}$ to the 
muon decay.
In our notation we split a generic bare coupling $g^{(0)}$ (the superscript referring 
to the order of the perturbative expansion) into the renormalized coupling $g$ and a 
counterterm $\delta g$:  $g^{(0)} = g + \delta g$. We decompose the VEVs and 
masses similarly, for example, $v^{(0)} = v + \delta v$ and $M^{(0)} = M_W + \delta M_W$. 

The Lagrangian \eqref{eq:lagrangian-gaugemass} containing the renormalized gauge boson 
masses can be recast as
\be 
\bsp
\cL_{\text{M}}^{\text{VB}} &= 
\bigl( M_W^2 + \delta M_W^2 \bigr)W_\mu^{+(0)}W^{-(0),\mu}
\\&
+\frac{1}{2}\bigl( M_Z^2 + \delta M_Z^2 \bigr) Z^{(0)}_\mu Z^{(0),\mu}
\\&
+\frac{1}{2}\bigl( M_{Z'}^2 + \delta M_{Z'}^2 \bigr) Z'^{(0)}_\mu Z'^{(0),\mu}
\\&
+ \bigl(\delta M_{ZA}^2 \Zb_\mu +\delta M_{Z'A}^2 \Zpb_\mu \bigr) A^{0,\mu}
\\&
+\delta M_{ZZ'}^2 \Zb_\mu {\Zpb}^\mu
\label{eq:lag-mass-barefield}
\esp
\ee
where $M_W$, $M_Z$ and $M_{Z'}$ are given in Eq.~\eqref{eq:vectormasses},
and the counterterms can be written symbolically as 
\be 
\bsp
\delta M_x^2 = 
\sum_{i=y,L,z,yz} c_{x,i} \delta g_{i}
+ c_{x,v} \delta v
+ c_{x,w} \delta w
\,,
\label{eq:massCT2}
\esp
\ee
$x=W$, $Z$, $Z'$, $ZA$, $Z'A$, $ZZ'$.
The coefficients $c_{x,i}$ are functions of the renormalized couplings 
$\gy$, $\gL$, $\gz$, $g_{yz}$, and VEVs $v$, $w$. In a similar way as done in the SM,
we can eliminate $\delta v$ in favor of the category (i) corrections,
\be 
\bsp
\delta \gL &= \frac{\delta e}{\sw} 
- \frac{e \cw^2}{ 2 M_W^2\sw^3}\biggl[\cw^2\biggl( 
\cz^2 \delta M_Z^2 + \sz^2 \delta M_{Z'}^2 
\\&
-  2\bigl(M_Z^2 - m^2_{Z'}\bigr)(\sz \delta \sz)
\biggr) -  \delta M_W^2 \biggr]
\label{eq:gl-counterterm}
\esp
\ee
where
\be 
\frac{2\delta e}{e} =  -\frac{\partial \Pi_{AA}(k^2)}{\partial k^2}\biggr|_{k^2 = 0} 
\!\!\!\!
- 2\frac{\sw}{\cw}\biggl( \cz\frac{ \Pi_{Z\!A}(0)}{M_Z^2} - \sz \frac{ \Pi_{Z'\!A}(0)}{M_{Z'}^2}\biggr),
\label{eq:delta-e}
\ee
and
\be 
\bsp
&\frac{\delta \sz}{\cz} = 
\frac{ \cz^2\Pi_{Z\!Z'}(M_{Z}^2)+\sz^2\Pi_{Z\!Z'}(M_{Z'}^2)}{M_Z^2-M_{Z'}^2}
\\&
-\frac{1}{2}\sz\cz \biggl(\frac{\partial \Pi_{ZZ}(k^2)}{\partial k^2}\biggr|_{k^2 = M_Z^2}
\!\!\!\!\! 
- \frac{\partial \Pi_{Z'\!Z'}(k^2)}{\partial k^2}\biggr|_{k^2 = M_{Z'}^2}\biggr)
\,.
\label{eq:delta-sz}
\esp
\ee
In Eq.~\eqref{eq:delta-sz} $\Pi$ is $(-\ri)$ times the transverse part of self energy graphs.
The one-loop charge renormalization counterterm $\delta e$ is exactly equal to the one-loop 
charge renormalization expression in the SM because the formula in the parenthesis is 
independent of $\theta_Z$. In the counterterm $\delta \sz$ the $Z-Z'$ mixing self energy 
$\Pi_{ZZ'}(M_{Z'}^2)$ and the derivatives $\partial_{k^2}\Pi_{ZZ}(k^2)$, 
$\partial_{k^2}\Pi_{Z'\!Z'}(k^2)$ appear as completely new contributions of the 
extended gauge sector through the renormalization of the $Z-Z'$ mixing angle $\theta_Z$.
We shall present the detailed derivation of our formulas for $\delta \gL$, $\delta e$ and
$\delta \sz$ elsewhere \footnote{Z.~P\'eli and Z.~Tr\'ocs\'anyi, in preparation.}.

Our main result is the complete one-loop prediction in U(1)$_z$ extensions of the SM 
to the parameter $\Delta r$ defined in Eq.~\eqref{eq:deltar-def}. 
In the on-shell renormalization scheme
\begin{align}
\Delta r_{\rm BSM} &= \frac{\text{Re}\Pi_{WW}(M_W^2)-\Pi_{WW}(0)}{M_W^2}
+\delta_{\text{BV}}
+\frac{2 \delta e}{e}
\nonumber\\&
+\frac{\cw^2}{\sw^2 M_W^2}\biggl[
\cw^2\text{Re}\Pi_{ZZ}(M_Z^2)
-\text{Re}\Pi_{WW}(M_W^2)
\biggr]
\nonumber\\&
-\sz^2\frac{\cw^2}{\sw^2}\frac{\cw^2}{M_W^2}\biggl[
\text{Re} \Pi_{ZZ}(M_Z^2)
- \text{Re} \Pi_{Z'\!Z'}(M_{Z'}^2)
\nonumber\\ &\qquad\qquad\qquad
+2\bigl(M_Z^2-M_{Z'}^2\bigr)\frac{\delta \sz}{\sz}
\biggr]
\label{eq:deltar-os}
\end{align}
where the first two lines are only formally the same as in the SM as they also include 
the BSM contributions. The one-loop self energies $\Pi_{WW}(k^2)$, $\Pi_{ZZ}(k^2)$ (category (ii) corrections) and the box and vertex contribution $\delta_{\text{BV}}$ 
(3rd category) have to be evaluated analogously to the SM, but with the inclusion of 
the BSM couplings and fields. We used the projection method of Ref.~\cite{Awramik:2002vu} 
to compute $\delta_{\text{BV}}$. The same applies to the last two terms where the new feature 
is that $\cw^2$ and $\sw^2 = 1- \cw^2$ has to be evaluated according to Eq.~\eqref{eq:mwmzmzp}. 

The expression \eqref{eq:deltar-os} must be finite and gauge independent as it collects 
the complete one-loop radiative corrections to the muon decay process. We checked 
explicitly that the $\epsilon$ poles cancel in $\Delta r$, and it is independent 
of the gauge parameters $\xi_{i}$, ($i=A$, $W$, $Z$, $Z'$).

To make numerical predictions, we adapted our findings to the \msbar renormalization 
scheme, employed frequently. We used the computational algorithm of Ref.~\cite{Athron:2022isz} 
where the prediction for the pole mass of the $W$ is expressed as
$M_W^2 = M_{W,{\rm SM}}^2 (1 + \Delta_W)$. In this equation we use the fit formula in 
Eq.~(45) of Ref.~\cite{Degrassi:2014sxa} for the 
standard model value $M_{W,{\rm SM}}^2$, while the correction term is written in terms of $\overline{\mathrm{MS}}$ renomalized parameters, denoted here by a hat,
\beq
\Delta_W = \frac{\hat{s}^2_{\rm W}}{\hat{c}^2_{\rm W}-\hat{s}^2_{\rm W}}
\left(\frac{\hat{c}^2_{\rm W}}{\hat{s}^2_{\rm W}} \Delta\hat{\rho} 
+ \Delta \hat{r}^{(1)}_{W}
\right)
\eeq
where $\hat{c}^2_{\rm W}$ and $\hat{s}^2_{\rm W}$ are the SM values computed as in 
Ref.~\cite{Degrassi:2014sxa}. As mentioned, the renormalization constant for the electric charge 
at one loop in the U(1)$_z$ extensions is exactly the same as in the SM, hence our formula for 
$\Delta_W$ does not contain the last term of Eq.~(5) in Ref.~\cite{Athron:2022isz}.

The term $\Delta \hat{\rho}$ is the difference of the full BSM and the SM 
predictions for $\hat{\rho}$, with formal expansion in perturbation theory at one loop accuracy as
$\Delta \hat{\rho} = \Delta \hat{\rho}^{(0)} + \Delta \hat{\rho}^{(1)}$ where at tree level
\beq
\Delta \hat{\rho}^{(0)} = \left(\frac{M_{Z'}^2}{M_Z^2} - 1\right) \sz^2
\,.
\eeq
Denoting the one-loop contributions in the 
full BSM by $\Delta \hat{\rho}^{(1)}_{\rm BSM}$, we have
\be
\bsp
&\Delta \rho^{(1)}_{\rm BSM} = \frac{1}{M_W^2}\biggl\{\text{Re}\Pi_{WW}(M_W^2)
\\&
    -\hat{c}^2_{\rm W}\biggl[\cz^2\biggl(\text{Re}\Pi_{ZZ}(M_Z^2) -2\sz\cz\Pi_{Z\!Z'}(M_Z^2) \biggr)
\\&\qquad
    + \sz^2 \biggl(\text{Re} \Pi_{Z'\!Z'}(M_{Z'}^2)
    -2\sz\cz \Pi_{Z\!Z'}(M_{Z'}^2)\biggr)
\\&\quad\quad
     + {\rm s}_Z^2 {\rm c}_Z^2
     \bigl(M_Z^2 - M_{Z'}^2\bigr)
\\&\quad\quad\times
\biggl(\frac{\partial \Pi_{ZZ}(k^2)}{\partial k^2}%\biggr|_{k^2 = M_Z^2}
\!\!\!\!\!
- \frac{\partial \Pi_{Z'\!Z'}(k^2)}{\partial k^2} \biggr)\biggr|_{k^2 = M_{Z'}^2}
    \biggr]
\biggr\}\,,
\label{eq:y1_msbar}
\esp
\ee
so the difference to the SM is 
\beq
\Delta \hat{\rho}^{(1)} = \Delta \hat{\rho}^{(1)}_{\rm BSM} - \Delta \hat{\rho}^{(1)}_{\rm SM}
\,.
\eeq
The term $\Delta \hat{r}^{(1)}_W$ collects the one-loop diagrammatic corrections to 
the muon decay process,
\be
\Delta \hat{r}^{(1)}_{W,{\rm BSM}}  =
\frac{\text{Re}\Pi_{WW}(M_W^2)- \Pi_{WW}(0)}{M_W^2} 
+\delta_{\text{BV}}
\,,
\label{eq:Deltar}
\ee
so the difference to the SM is 
\beq
\Delta \hat{r}^{(1)}_W = \Delta \hat{r}^{(1)}_{W,{\rm BSM}} - \Delta \hat{r}^{(1)}_{W,{\rm SM}}
\eeq
where the subtracted term is formally the same as formula \eqref{eq:Deltar}, but computed with SM degrees of freedom.

The mass $M_W$ can also be computed with automated programs once the model and the 
input parameters are defined, see for instance, \texttt{SARAH/SPheno}
\cite{Porod:2003um,Porod:2011nf,Staub:2009bi,Staub:2013tta} and \texttt{FlexibleSusy} 
\cite{Athron:2014yba,Athron:2022isz}. However, the predictions for 
$M_W$ in U(1) extensions provided by these programs employ approximate one-loop BSM 
corrections for $\Delta \hat{\rho}^{(1)}_{\rm BSM}$ in U(1)$_z$ type extensions. Our goal here is to check and explore by a numerical analysis the validity 
of these approximations depending on the values of the input parameters, and 
to point out that such automated computations can lead to significantly 
different prediction than ours.

We investigate the predictions for $M_W$ at fixed renormalization scale $\mu = M_Z$ in
the \msbar scheme in two approximations: 
(i) one includes the complete set of one-loop radiative corrections and two-loop SM corrections computed by us,
(ii) a truncation when
the one-loop radiative corrections to $\hat{\rho}$ are formally the same as in the SM
\be
  \Delta \rho^{(1)}_{\rm BSM} = \frac{1}{M_W^2}\biggl[\text{Re}\Pi_{WW}(M_W^2)
    -\hat{c}^2_{\rm W}\text{Re}\Pi_{ZZ}(M_Z^2)\biggr]\,,
\ee
but with self-energies evaluated in the BSM extension, which is the U(1)$_z$ extension 
in our work. Case (ii) is implemented in automated high energy physics tools 
such as \texttt{SARAH/SPheno} \cite{Porod:2003um,Porod:2011nf,Staub:2009bi,Staub:2013tta} 
and \texttt{FlexibleSusy} \cite{Athron:2014yba,Athron:2022isz}.

We use the set of input parameters
\begin{equation*}
\bsp
\GF = 1.1663787 \cdot 10^{-5}\,\GeV^{-2},
\;
M_Z= 91.1876\,\GeV\,,
\\
M_H = 125.25\,\GeV,
\; 
m_t = 172.83\,\GeV,
\;
m_b = 4.18\,\GeV\,,
\\
\alpha_s(M_Z) = 0.1179,
\;
\alpha = \bigl(137.036\bigr)^{-1},
\;
\Delta\alpha_{\rm had}^{(5)} = 0.02760
\esp
\label{eq:numinput}
\end{equation*}
taken from \cite{Workman:2022ynf}. In particular, the value of the top quark pole mass $m_t$ 
is presented in the Quark Masses subsection of Chapter 10 and the numerical value for 
$\Delta\alpha_{\rm had}^{(5)}$ is the average of the results presented in Table~10.1 of 
Ref.~\cite{Workman:2022ynf}.

Once the parameters in \eqref{eq:numinput} are set, the prediction for $M_W$ at fixed $\mu$ depends on five free parameters
$M_{Z'},~\sz,~M_S,~\sss,~\tanb$
where $M_S$ is the mass of the scalar particle appearing after SSB of the complex scalar field $\chi$
and $\sss$ is the scalar mixing angle. The SM is recovered in the limit of vanishing massive neutral gauge boson and scalar
mixings, $\sz = \sss = 0$, which produces our reference SM predictions in agreement with the literature,
\begin{equation*}
\bsp
M_{W,{\rm SM}} = 80.353\,\GeV,~~\hat{s}^2_{{\rm W,SM}}(M_Z) = 0.2313,
\\
\hat{\alpha}^{-1}_{\rm SM}(M_Z) = 127.952\,,
\qquad\qquad
\esp
\end{equation*}
once the decoupling of the top-quark \cite{Fanchiotti:1992tu} is applied. 

The extension of the SM gauge sector affects the 
vector and axial vector (V-A) couplings of the $Z$ boson and introduces the $Z'$ boson, 
which also interacts with fermions through its own V-A couplings. The exact form of these 
couplings depend on the $z$-charge assignment of the new U(1)$_z$ gauge group. 
In order to present numerical values for our predictions, we select the SWSM
where the $z$ charges are fixed as $z_Q = 1/6$, $z_U = 7/6$ and 
$z_\phi = z_U-z_Q$~\cite{Trocsanyi:2018bkm}.

We compare the predictions of the two cases in order to explore the validity 
of the approximations applied in (ii). We present our findings as benchmark points 
expressed as the differences
$\Delta M_W = M_W - M_{W,{\rm SM}}$,
sampled from different regions of the parameter space spanned by $M_{Z'},~\sz,~M_S,~\sss$
and $\tanb$. In general, the mixing angle $\theta_{\rm Z}$ severely affects the predictions 
for electroweak observables. We present our benchmark points depending on the ratios 
$M_{Z'}/M_Z$ and $M_S/M_h$ being smaller or  larger than one.

\begin{figure}[t]
\includegraphics[width=0.8\linewidth]{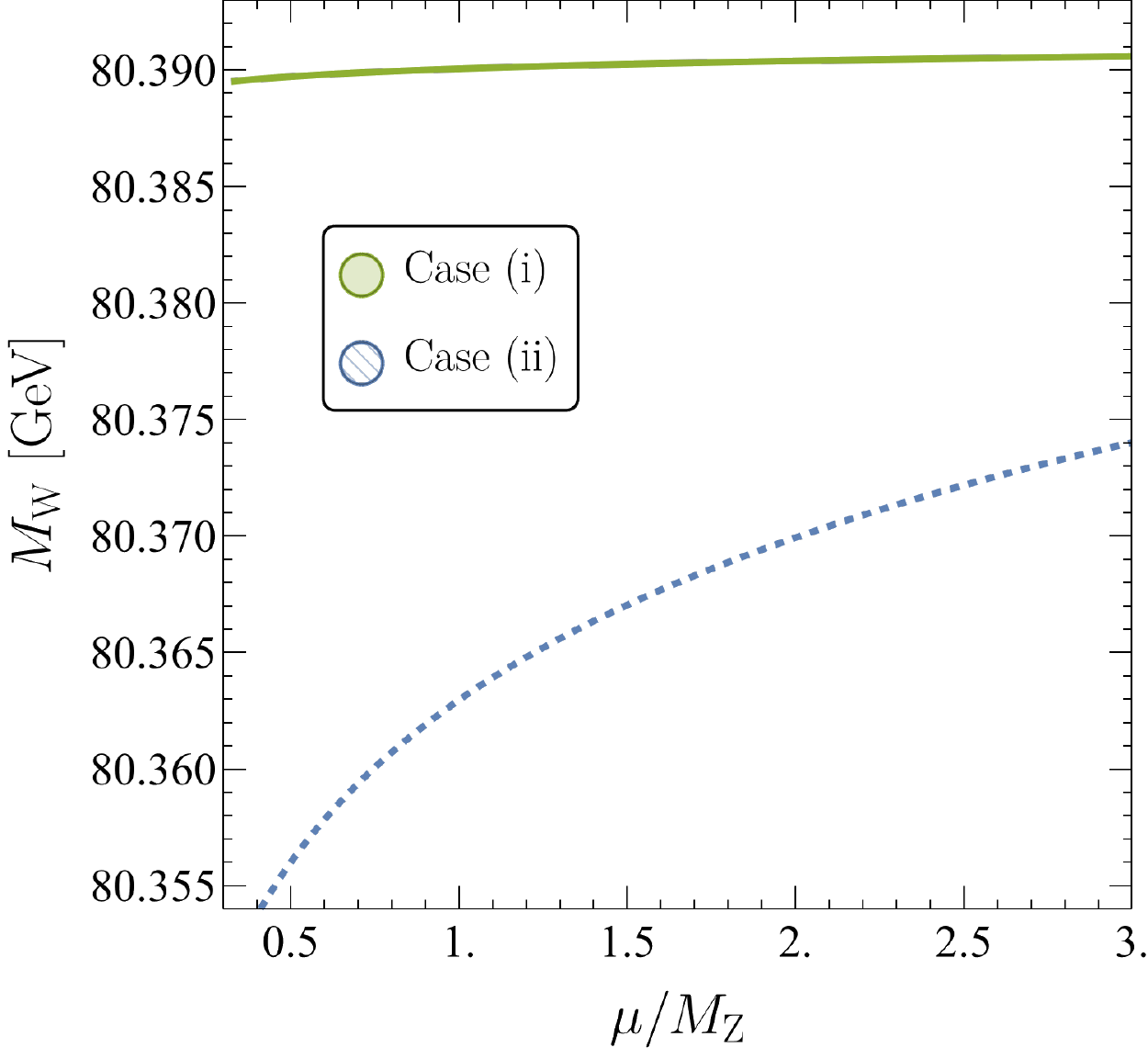}
\caption{\label{fig:scaledep} 
The dependence of our prediction for $M_W$ on the renormalization scale $\mu$ in case (i) (solid, green curve) and (ii) (dotted blue curve).}
\end{figure}

A light (heavy) $Z'$ boson  with $M_{Z'}<M_{Z}$ ($M_{Z'}>M_{Z}$)  contributes a negative 
(positive) shift to the mass of the $W$ boson. For light new physics (Tab.~\ref{table:p0b}), $M_{Z'}/M_{Z}\ll 1$, both cases
provide a good approximation. For heavy new physics (Tab.~\ref{table:p2b}) however, 
when $M_{Z'}/M_{Z}\gg 1$, the approximation of case (ii) may become unreliable 
and the difference from the complete prediction (i) can surpass the size of the typical experimental uncertainty of about $10\,\MeV$.
\begin{table}[t!]
\centering
\begin{tabular}{|l || c | c || c | c |} 
\hline
\hline
 $~\quad s_Z$ & \multicolumn{4}{|c|}{$5\cdot 10^{-4}$} \\
\hline
\backslashbox{$\tanb\!\!\!\!$}{$\!\!\!\!M_S$} 
& \multicolumn{2}{|c||}{$\begin{array}{c}0.5\,\TeV\\\text{(i)~~~(ii)}\end{array}$}
& \multicolumn{2}{|c|}{$\begin{array}{c}5\,\TeV\\\text{(i)~~~(ii)}\end{array}$} 
\\
\hline
\hline
 ~~0.1 & --1 & --1 & --2 & --2 \\ 
 \hline
 ~~1 & --1 & --1 & --2 & --2 \\ 
 \hline
 \,10 & --1 & --1 & --2 & --2 \\ 
 \hline
 \hline
\end{tabular}
\caption{Predictions for $\Delta M_W = M_W - M_{W,{\rm SM}}$ in MeV units at parameter values $M_{Z'}=50\,\MeV$, $\sz = 0.005$ and $\sss=0.1$.}
\label{table:p0b}
\end{table}
%%
\iffalse
\begin{table}[t!]
\centering
\begin{tabular}{|l || c | c || c | c |} 
\hline
%   &  \multicolumn{2}{| c ||}{$\Delta M_W$\,[\MeV]} & \multicolumn{2}{| c ||}{$\Delta M_W$\,[\MeV]} \bigstrut\\
\hline
\backslashbox{$\tanb$}{$M_S$} & \multicolumn{2}{|c||}{ 50\,\GeV}
& \multicolumn{2}{|c|}{500\,\GeV} \bigstrut\\
\hline
\hline
%~~ & (i) & (ii) & (i) & (ii) \\
%\hline
% ~~2 & --164 & --736 & 32 & --739 \\ 
%\hline
% ~~5 & 4 & --82 & 34 & --83 \\ 
%\hline
10 & 31 & 11 & 37 & 10 \\
\hline 
20 & 39 & 35 & 39 & 34 \\ 
\hline
30 & 41 & 40 & 40 & 38 \\ 
\hline
\end{tabular}
\caption{Predictions for $\Delta M_W = M_W - M_{W,{\rm SM}}$ in MeV units at parameter values $M_{Z'}=5~\TeV$, $\sz = 5\times 10^{-4}$ and $\sss=0.1$.}
\label{table:p2b}
\end{table}
%%
\fi
%%
\begin{table}[t!]
\centering
\begin{tabular}{|l || c | c || c | c || c | c || c | c |} 
\hline
\hline
 $~\quad s_Z$ & \multicolumn{4}{|c||}{$5\cdot 10^{-4}$} & \multicolumn{4}{|c|}{$7\cdot 10^{-4}$}\\
\hline
\backslashbox{$\tanb\!\!\!\!$}{$\!\!\!\!M_S$} 
& \multicolumn{2}{|c||}{$\begin{array}{c}0.5\,\TeV\\\text{(i)~~~(ii)}\end{array}$}
& \multicolumn{2}{|c||}{$\begin{array}{c}5\,\TeV\\\text{(i)~~~(ii)}\end{array}$} 
& \multicolumn{2}{|c||}{$\begin{array}{c}0.5\,\TeV\\\text{(i)~~~(ii)}\end{array}$}
& \multicolumn{2}{|c|}{$\begin{array}{c}5\,\TeV\\\text{(i)~~~(ii)}\end{array}$}
\\
\hline
\hline
10 & ~37~ & 10 & ~35~ & 13 & ~75~ & 29 & ~73~ & 36 \\
\hline 
20 & ~39~ & 34 & ~35~ & 34 & ~81~ & 76 & ~74~ & 79 \\ 
\hline
30 & ~40~ & 38 & ~35~ & 37 & ~83~ & 85 & ~75~ & 85\\ 
\hline
\hline
\end{tabular}
\caption{Predictions for $\Delta M_W = M_W - M_{W,{\rm SM}}$ in MeV units at parameter values $M_{Z'}=5~\TeV$ and $\sss=0.1$.}
\label{table:p2b}
\end{table}
%%
\iffalse
%%
\begin{table}[t!]
\centering
\begin{tabular}{|l || c | c || c | c |} 
\hline
%   &  \multicolumn{2}{| c ||}{$\Delta M_W$\,[\MeV]} & \multicolumn{2}{| c ||}{$\Delta M_W$\,[\MeV]} \bigstrut\\
\hline
\backslashbox{$\tanb$}{$M_S$} & \multicolumn{2}{|c||}{ 500\,\GeV}
& \multicolumn{2}{|c|}{5000\,\GeV} \bigstrut\\
\hline
\hline
~~ & (i) & (ii) & (i) & (ii) \\
\hline
10 & 165 & 99 & 163 & 117 \\
\hline 
20 & 182 & 196 & 170 & 203 \\ 
\hline
30 & 187 & 214 & 173 & 216 \\ 
\hline
\end{tabular}
\caption{Predictions for $\Delta M_W = M_W - M_{W,{\rm SM}}$ in MeV units at parameter values $M_{Z'}=5~\TeV$, $\sz = 10^{-3}$ and $\sss=0.1$.}
\label{table:p2b}
\end{table}
%%
% CDFII COMPATIBLE VALUES BELOW
%%
\begin{table}[t!]
\centering
\begin{tabular}{|l || c | c || c | c |} 
\hline
%   &  \multicolumn{2}{| c ||}{$\Delta M_W$\,[\MeV]} & \multicolumn{2}{| c ||}{$\Delta M_W$\,[\MeV]} \bigstrut\\
\hline
\backslashbox{$\tanb$}{$M_S$} & \multicolumn{2}{|c||}{ 500\,\GeV}
& \multicolumn{2}{|c|}{5000\,\GeV} \bigstrut\\
\hline
\hline
~~ & (i) & (ii) & (i) & (ii) \\
\hline
10 & 75 & 29 & 73 & 36 \\
\hline 
20 & 81 & 76 & 74 & 79 \\ 
\hline
30 & 83 & 85 & 75 & 85 \\ 
\hline
\end{tabular}
\caption{Predictions for $\Delta M_W = M_W - M_{W,{\rm SM}}$ in MeV units at parameter values $M_{Z'}=5~\TeV$, $\sz = 7 \times 10^{-4}$ and $\sss=0.1$.}
\label{table:p2b}
\end{table}
%%
\fi
%%
The advantage of the computational algorithm of Ref.~\cite{Athron:2022isz} 
is that it removes the non-decoupling logarithmic contributions from the one-loop 
formula. Those logarithms can become potentially large and are canceled in 
perturbation theory only if the two-loop BSM contributions are also computed. The 
cancellation of those large logarithms can be seen in Fig.~\ref{fig:scaledep} where we
present the dependence of our prediction for $M_W$ on the renormalization scale $\mu$. 
The solid line corresponds to case (i), while the dotted one to case (ii). 
The input values are the same as in Tab.~\ref{table:p2b} with $\tan\beta = 10$ and 
$M_S = 500\,\GeV$.

\begin{figure}[t]
{
\includegraphics[width=0.8\linewidth]{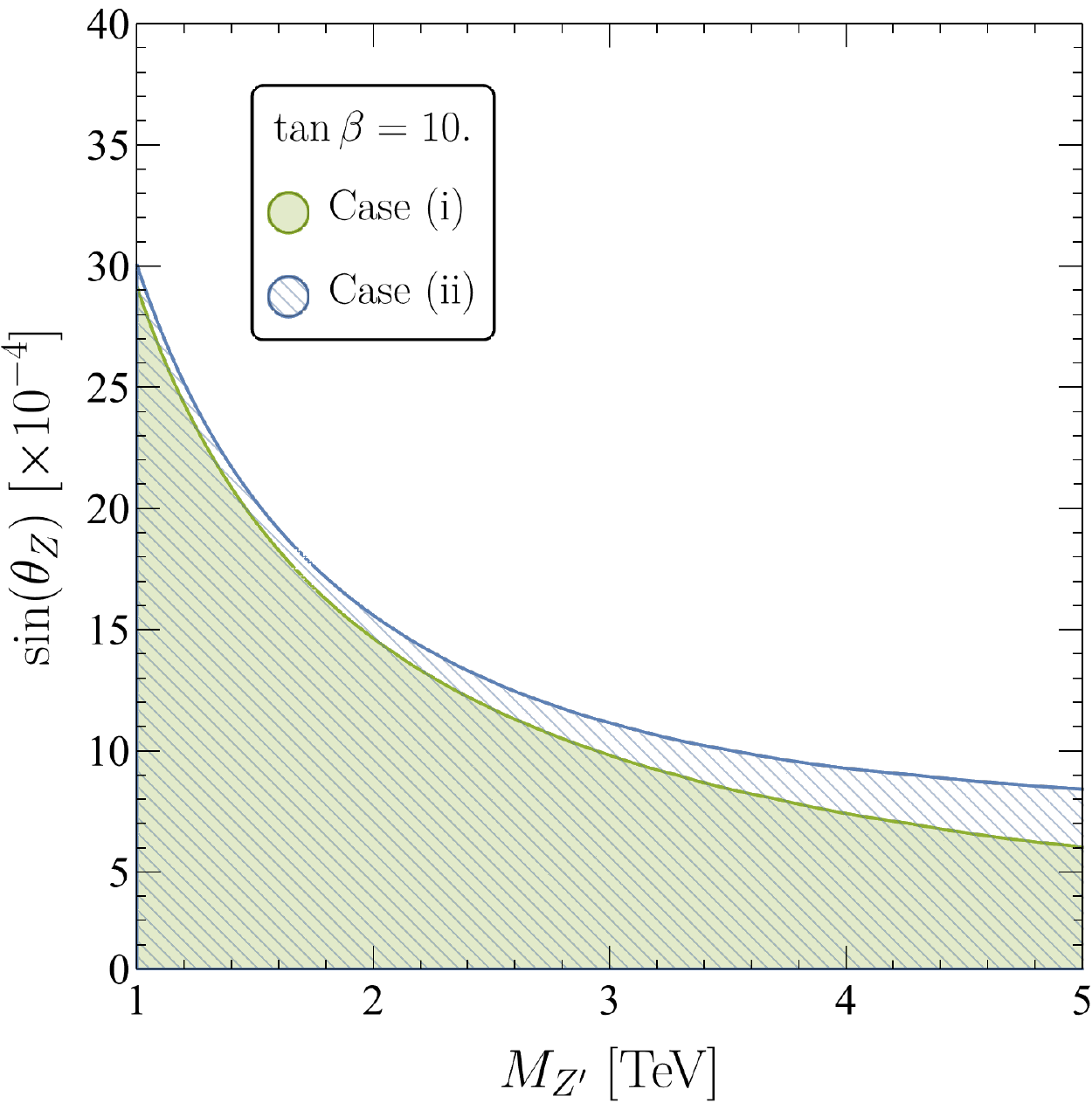}
}
\caption{\label{fig:2} 
Comparison of the predictions by plotting the regions allowed by the requirement 
$|M_W^{\rm exp.}-M_W| < 2\sigma$ for the fixed values 
$\sss=0.1$ and $M_S=500\,\GeV$ in the $M_{Z'}-\sz$ plane for large values of $M_{Z'}$.
}
\end{figure}

We also compare the two predictions by showing the $2\sigma$ allowed band 
$|M_W^{\rm exp.} - M_W| < 2 \sigma$ where $M_W$ is our theoretical prediction 
in the U(1)$_z$ extension, $M_W^{\rm exp.}$ is the experimentally measured value and 
$\sigma = \sqrt{\sigma_{\rm exp.}^2 + \sigma_{\rm theo.}^2 + \sigma_{\rm param.}^2}$
with $\sigma_{\rm exp.}$ being the uncertainty of $M_W^{\rm exp.}$, $\sigma_{\rm theo.}$ 
being the theoretical and $\sigma_{\rm param.}$ being the parametric uncertainty of our 
prediction $M_W$. The theoretical uncertainty is estimated in Ref.~\cite{Degrassi:2014sxa} 
to be $\sigma_{\rm theo.}=4\,\MeV$, while we estimate the parametric uncertainty with 
the input values presented in Eq.~\eqref{eq:numinput} to be $\sigma_{\rm theo.}=8\,\MeV$.
Figure~\ref{fig:2} shows the $2\sigma$ allowed bands obtained with the PDG world average 
\cite{Workman:2022ynf} for $M_W^{\rm exp.}$. We see that the approximation (ii) leads to 
different allowed regions for a heavy $Z'$ whose extent depends on the values of the 
free parameters. This warns us that one has to be careful when using the automated 
computations for the radiative corrections to the mass of the $W$ boson.

In summary, we computed the one-loop corrections $\Delta r$ 
to the muon decay in a general U(1)$_z$ extension of the standard model. We did not make any
assumptions for the parameters of the model or employing any truncations beyond the systematic
perturbation theory. We presented the complete expression for $\Delta r$ in the on-shell
renormalization scheme and also in the \msbar scheme. 
We found that not only additional 
loops appear in the transverse $W$ and $Z$ boson self energies $\Pi_{WW}(p^2)$, $\Pi_{ZZ}(p^2)$ 
and in the box- and vertex corrections $\delta_{\rm BV}$ but the $Z'$ boson self energy 
$\Pi_{Z'\!Z'}(p^2)$ and the wave function renormalizations $Z_{ZZ}$, $Z_{Z'\!Z'}$ and $Z_{Z\!Z'}$ 
also contribute to $\Delta r$. The new terms appear in the renormalization of $\theta_W$ 
in the on-shell scheme or in the $\hat{\rho}$ parameter in the \msbar scheme.

The high energy physics tools that can automatically compute the radiative corrections to $M_W$
in BSM models neglect several terms from the complete expression in Eq.~\eqref{eq:y1_msbar} 
for U(1)$_z$ type BSM extensions. We pointed out using a specific example model, the SWSM
\cite{Trocsanyi:2018bkm} that the effect of the neglected terms can affect significantly the
prediction. In qualitatively different regions of the parameter space spanned by the free input
parameters, we selected benchmark points for small $Z-Z'$ mixing $\sz$. We found that neglecting
$Z_{ZZ}$, $Z_{Z'\!Z'}$ and $\Pi_{Z'\!Z'}(p^2)$ produces small $\mathcal{O}(\MeV)$ numerical 
differences for $\tanb \geq 1$ in the region where the mass of the new netural gauge boson is much
lighter then the $Z$ boson, but for $M_{Z'} \gg M_Z$ the use of the complete expression 
employed in case (i) is more appropriate.

\acknowledgments
We thank P.~Athron for useful 
correspondence on the proper treatment of non-decoupling logarithms.

%\bibliographystyle{apsrev4-1}
%\bibliography{swsm_ew.bib}
%

\end{document}